# THE GEOLOGICAL AND CLIMATOLOGICAL CASE FOR A WARMER AND WETTER EARLY MARS

Ramses M. Ramirez[i]

Robert A. Craddock[ii]

[i]Earth-Life Science Institute (ELSI), Tokyo Institute of Technology, Tokyo 152-8550, Japan (Formerly: Cornell Center of Astrophysics and Planetary Science, Cornell University, Ithaca, NY, USA, 14850)

[ii]Center for Earth and Planetary Studies, National Air and Space Museum, Smithsonian Institution, Washington, D.C., USA, 20560

**The climate of early Mars remains a topic of intense debate. Ancient terrains preserve landscapes consistent with stream channels, lake basins, and possibly even oceans, and thus the presence of liquid water flowing on the Martian surface 4 billion years ago. However, despite the geological evidence, determining how long climatic conditions supporting liquid water lasted remains uncertain. Climate models have struggled to generate sufficiently warm surface conditions given the faint young Sun - even assuming a denser early atmosphere. A warm climate could have potentially been sustained by supplementing atmospheric $CO_2$ and $H_2O$ warming with either secondary greenhouse gases or clouds. Alternatively, the Martian climate could have been predominantly cold and icy, with transient warming episodes triggered by meteoritic impacts, volcanic eruptions, methane bursts, or limit cycles. Here, we argue that a warm and semi-arid climate capable of producing rain is most consistent with the geological and climatological evidence.**

Today, the surface of Mars resembles that of a hyper-arid desert. Perennially dry and cold, mean surface temperatures are ~70 K lower than on the Earth, and the atmosphere is only ~1% as thick. Although recent evidence suggests that small amounts of water may flow seasonally[1], the Martian surface is still colder and drier than most locations on Earth. Although Martian surface temperatures near the equator can exceed the freezing point of water, mean surface temperatures are only ~218 K[2].

However, this description of Mars today is in stark contrast to what is recorded in the ~ 4-billion-year old (late Noachian to early Hesperian) landscape of the southern hemisphere. Here, ancient terrains reveal a very different planet – a once warmer and wetter climate preserved in a wide array of surface fluvial features[3]. These include lakes, channels, modified craters, alluvial fans, and deltas[4-6]. On a grander scale, possible ancient shorelines suggest an early ocean may have once covered much of the northern hemisphere[7], leading to estimates that the early Martian ocean could have been equivalent to a global water body ~550 meters deep[8]. Subsequent studies using hydrogen isotope ratios of known water reservoirs computed a smaller initial water inventory 4.5 billion years ago of about 137 to 164 meters deep[9]. This is a minimum estimate of the initial inventory, however, as both gases could have escaped indiscriminately when the Sun was very young and much more active[9].

The widespread occurrence of clays also strongly supports the notion of persistent water on ancient Mars[10,11]. These fine-grained materials form from the gradual weathering of basaltic rocks exposed to water.

However, among the most compelling evidence of a drastically different past climate is the widespread presence of valley networks. The ones created at the Noachian/Hesperian transition (~3.8 – 3.6 Ga) often extend hundreds – or even thousands – of kilometers and can be tens to hundreds of meters deep, easily rivaling some of the largest erosional features on Earth (Figure 1). Nevertheless, this does not mean that early Mars was globally as warm and wet as modern Earth. Image analyses suggest that Martian valleys, located in the southern cratered highlands (Figure 2), may be less developed than terrestrial systems[5,12], and although it is generally understood that the Martian valleys were also generated by rainfall and surface runoff, such conditions did not last long enough for valley networks to fully integrate with the cratered landscape.[4-6,12,13] Admittedly, such comparisons are imperfect because younger systems on Earth are better preserved[3,14] and benefit from much higher resolution data sets[15].

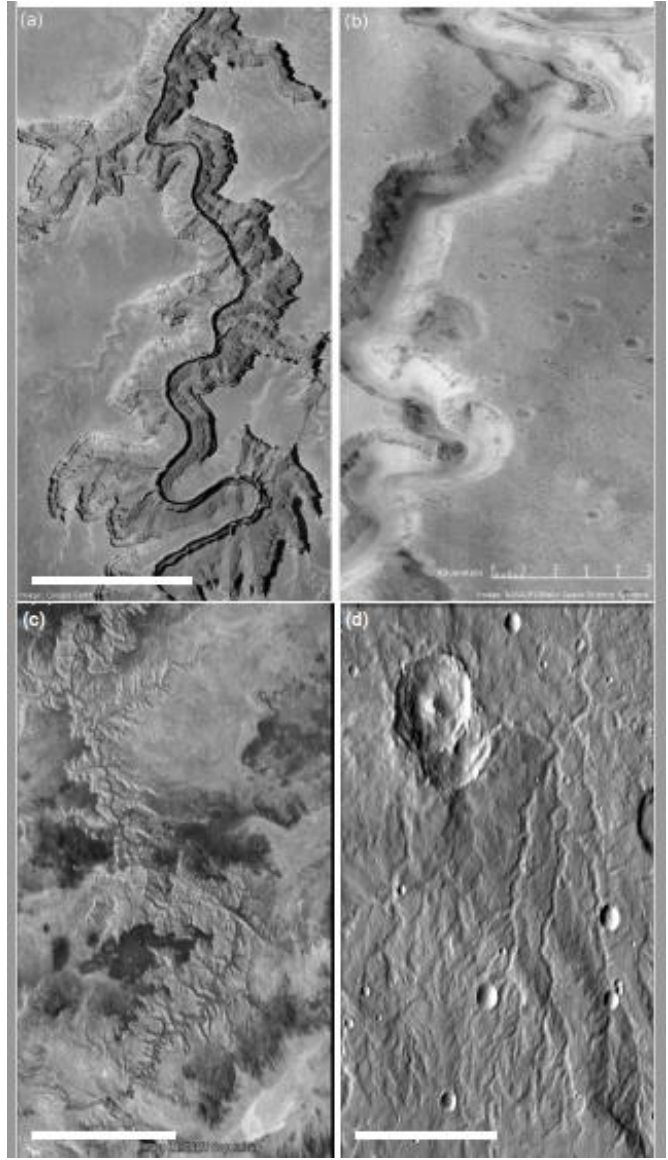

**Figure 1:** Geomorphic evidence for water on ancient Mars compared to Earth. (a) section of the Colorado River Canyon is compared against (b) part of Mars's Nanedi Vallis (Lunae Palus quadrangle, 4.9°N,49°W). A river had once cut through Nanedi Vallis (top of image), which continues for over 500 km (not shown). (Image Credit: Chester Harman). (c) The Grand Canyon versus (d) a Martian dendritic river system (Arabia Quadrangle, 12°N,43°E.). The white scale bars are 5km (top panel) and 60 km long (bottom panel), respectively. Slight morphologic differences between terrestrial and Martian comparisons are attributable to the great differences in age.

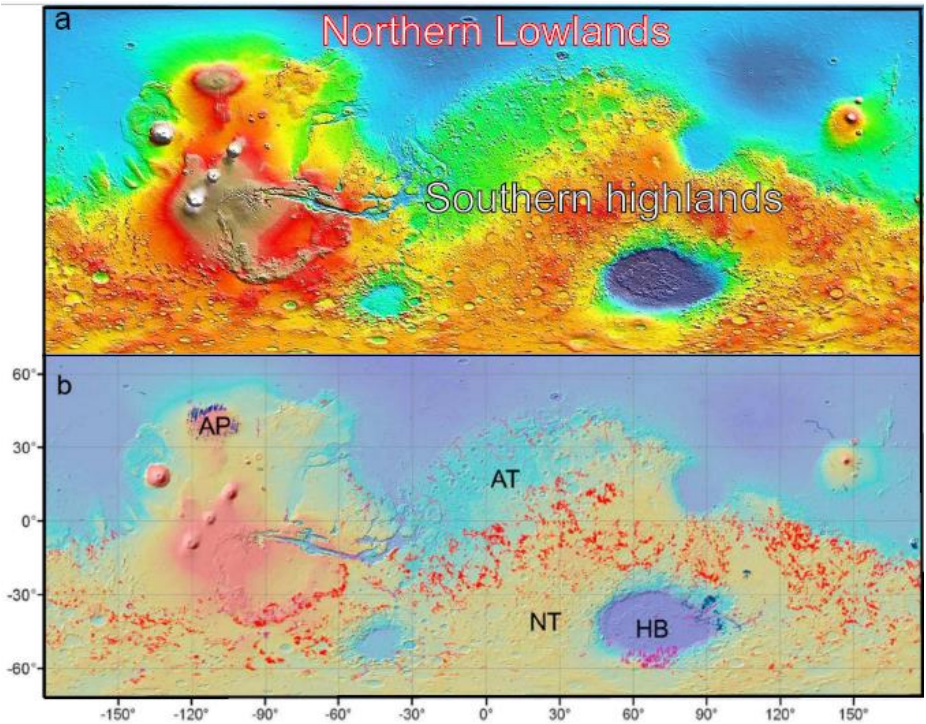

**Figure 2:** Marian valley networks. Mars Orbiter Laser Altimeter elevation map (a) with high (red) and low (blue) elevations and (b) valley network distribution. THEMIS data were used over a MOLA shaded relief map with Amazonian (cyan), Hesperian (purple), and Noachian (orange-red) terrains. The highest concentration of valleys is on Noachian terrains near the equator. (AP, Alba Patera; AT, Arabia Terra; HB, Hella Basin, NT; Noachis Terra) (as mapped by Hynek et al.[15]).

In spite of the geologic evidence, attaining warmer and wetter past conditions on early Mars has proven exceedingly difficult to model. Mars is located ~1.5 times farther from the Sun than the Earth is and receives less than half the solar energy that our planet does. Moreover, the Sun was ~25% dimmer and the solar energy input was only ~1/3 that received by the modern Earth on *early* Mars. This presents daunting challenges for explaining a potentially warm early Mars.

Initially, a dense $CO_2$-rich atmosphere with $CO_2$ partial pressures ranging from about 2 - 10 bars was invoked for early Mars[16]. On face value, such high $CO_2$ pressures are not implausible. Earth and Venus were endowed each with ~60 – 90 bars of $CO_2$[17]. If Mars had received the same amount of carbon per unit mass, its initial $CO_2$ inventory could have been ~ 8 – 13 bars, after adjusting for gravity due to its smaller size. However, the young Sun was extraordinarily active during this time and intense EUV fluxes would have removed much, if not all, of that initial $CO_2$[18]. Alternatively, some of this carbon may have been sequestered within the rocks, possibly supplementing volcanic outgassing of $CO_2$ later on if Mars had plate tectonics[19], but this remains controversial[20].

Regardless, the highest surface temperatures and pressures achieved by models in a putative $CO_2$-$H_2O$ atmosphere is ~230 K and ~2 - 3 bar of $CO_2$[21-24], which is insufficient to support warm, wet conditions (Supp. Info. Figure S1). Two primary problems persist as to why a $CO_2$-$H_2O$ greenhouse effect is ineffective at warming early Mars. First, $CO_2$ condensation intensifies as pressure increases, which weakens the greenhouse effect and lowers the mean surface temperature. In 3-D models, this results in atmospheric collapse[22]. Secondly, $CO_2$ scatters sunlight ~2.5 times more efficiently than the Earth's $N_2$-rich atmosphere, and at high enough pressures, the amount of sunlight reflected to space is greater than that absorbed, causing net cooling[25].

In spite of the difficulties in achieving sufficient warming in climate simulations, it is difficult to explain the geological evidence with permanently cold conditions on the early Martian surface. Instead, a transiently warm -if not continuously warm- climate seems to be required. With this framework in mind, we summarize some recent ideas to model a warm early Mars. We then assess the geologic and climatological evidence and discuss whether it is consistent with recent hypotheses that assume a cold and icy baseline planet that becomes transiently warm[26]. We conclude with our preferred interpretation and discuss limitations and outstanding questions (see Figure 3 for an events timeline).

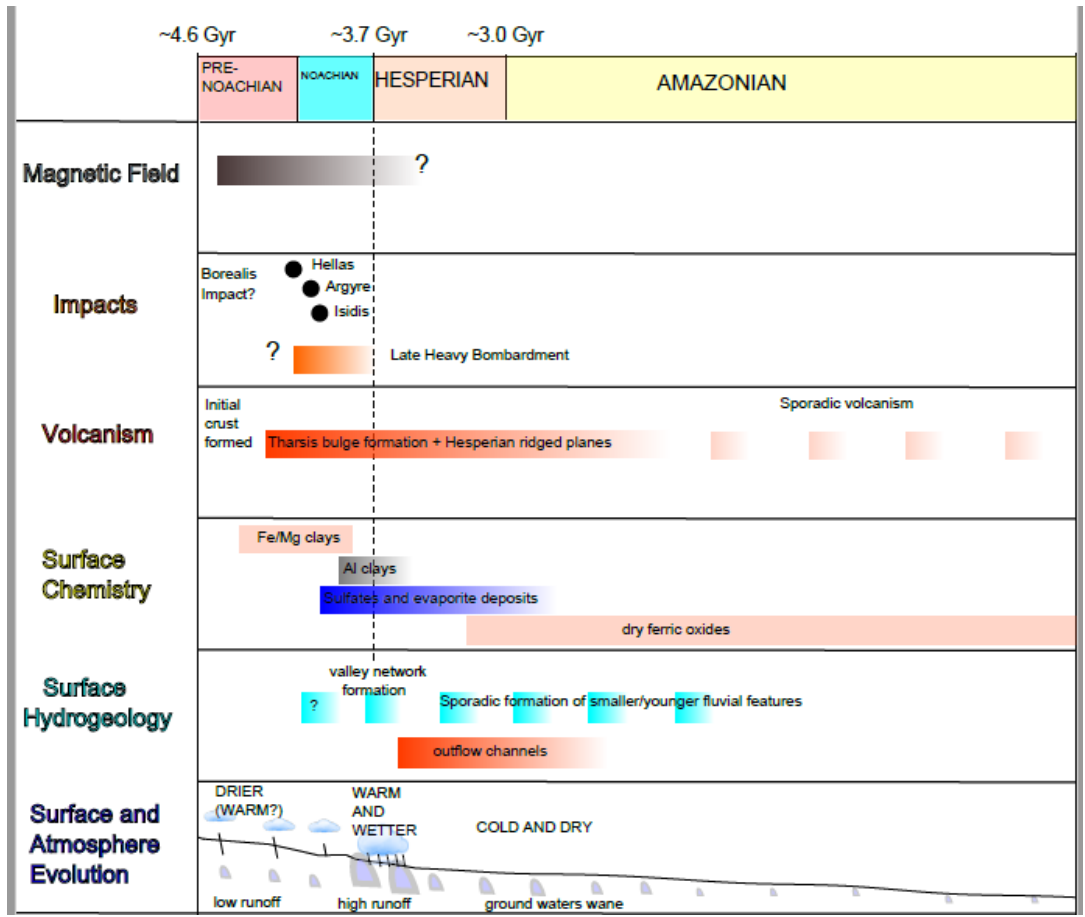

**Figure 3**: A schematic of Mars' geologic evolution with time. We propose that early Mars started with a dry and possibly warm climate that triggered valley network incision once conditions for surface runoff at the Noachian/Hesperian transition became favorable enough to support fluvial erosion[3,5-6,79]. Question marks refer to uncertainties in timespans. Based on data and analyses from refs 3, 62.

## WARMING EARLY MARS

The addition of $CO_2$ clouds could enhance the surface warming in a dense $CO_2$ atmosphere[27]. Such clouds, if composed of sufficiently large (10 – 100-μm) particles, could have backscattered enough infrared radiation to warm the surface tens of degrees and generate above-freezing mean surface temperatures assuming 100% cloud cover[27]. The warming is greatly diminished at more realistic cloud fractions[22] and, moreover, $CO_2$ clouds actually cool the climate if located at sufficiently low altitudes[28,29].

Given the difficulties that arise in producing warm climates with just $CO_2$ and $H_2O$, it has been suggested that other secondary atmospheric greenhouse gases could have played a key role. The common volcanic gas $SO_2$ is intriguing given the evidence of widespread volcanism and the occurrence of sulfur minerals in ancient terrains[10]. Indeed, several studies have shown that this secondary greenhouse gas can produce significant warming[30-32]. However, $SO_2$ forms highly reflective sulfate and sulfur particles (i.e. aerosols) that help dampen the greenhouse effect and is short-lived, raining out of the atmosphere once warm conditions are achieved[21]. Furthermore, global climate model (GCM) simulations suggest that any transient warming conditions produced by $SO_2$ do not generally coincide with the locations of the valley networks[33].

Another potential secondary greenhouse gas is $CH_4$, which would outgas on an early Mars if the mantle was oxygen-poor[23]. Originally, it was thought the $CH_4$ greenhouse effect was weak because warming in the lower atmospheric would be negated by warming in the upper atmosphere[23]. However, collisions between $CO_2$ and $CH_4$ molecules in a dense $CO_2$ atmosphere could significantly enhance the $CH_4$ greenhouse effect[24]. A 2-bar $CO_2$ atmosphere with $CH_4$ ratios exceeding ~10% can raise mean surface temperatures close to the freezing point of water. Such high $CH_4$ concentrations appear difficult to maintain, however. At $CH_4/CO_2$ ratios above ~0.1, anti-greenhouse hazes form, cooling the climate[34,35]. Moreover, $CH_4$ can react with $H_2O$ vapor, forming $CO_2$ instead[36].

Hydrogen, $H_2$, outgassed by volcanoes could have provided the additional warming[23,24,37]. Although hydrogen is normally a poor greenhouse gas, it becomes effective in collisions with a dense background gas. In this way, absorption by $H_2$ is even more effective than that for $CH_4$. This is because 1) absorption by $CO_2$-$H_2$ is stronger than that for $CO_2$-$CH_4$[24]; 2) $CO_2$-$H_2$ absorbs in spectral regions where $CO_2$ and $H_2O$ absorb poorly[23]; and 3) absorption by $CH_4$ is significant at solar wavelengths[23]. Simulations suggest that sufficient $CO_2$-$H_2$ warming is possible (Box 1) at $CO_2$ pressures low enough to be consistent with the scarcity of surface carbonates identified on Mars[37] (Box 1). $H_2$ concentrations exceeding a couple of percent satisfy atmospheric pressure constraints (< ~2 bar)[24,37,38,39]. Nevertheless, the pressures required to achieve warm solutions can be significantly higher depending on the surface ice fraction (discussed below)[37]. Although hydrogen concentrations >~5% may have been hard to achieve, they could have been possible with high outgassing rates or at escape rates below the diffusion limit[36]. A $CO_2$-$CH_4$ greenhouse also works better when enhanced with $CO_2$-$H_2$[24].

# BOX 1

*THE CARBONATE PROBLEM*

The so-called "carbonate problem" is sometimes invoked to argue against a warmer and wetter early Mars, because a thicker $CO_2$ –rich atmosphere should have resulted in the formation of widespread surface carbonates[83]. However, only trace amounts of carbonates have ever been detected from orbit[84]. The problem seems even more challenging given that atmospheric escape processes over the last 4 billion years may not have removed more than ~0.1 bar $CO_2$[85]. Nevertheless, the lack of observed carbonates in ancient terrains is not inconsistent with a warmer and wetter climate. The resultant pH of rainwater in a dense and warm $CO_2$ atmosphere could be ~5 or less, which is acidic enough to dissolve surface carbonates and transport them into the subsurface[78]. The possibility that such a process occurred is supported by carbonate-bearing Martian meteorities[86], suggesting a crustal sink for carbonates exists. Surface carbonates may also be obscured by recent soils, subsequently altered, covered by dust, or have evaded detection via blocky surfaces or insufficient spectral resolution[3]. Because carbonates should be forming in the current Martian environment, some process must be causing their masking or destruction[3,87]. Recent climate modeling results[37] found warm solutions with total pressures well below 1 bar, which satisfies prehnite stability[38]. Thus, relatively thin atmospheres do not necessarily require that early Mars was cold.

*THE PRESENCE OF OLIVINE AND OTHER MINERALS ON MARS*

The mineralogy of the Martian surface has been used as an argument against a warm early Mars. For example, olivine weathers quickly in the presence of water, so its presence on Martian terrains has led to suggestions that past Martian climates could not have been warm and wet[88]. However, many olivine deposits are located in volcanic materials that are much younger than the valley networks[89]. Similarly, the limited chemical alteration found in Gale crater by the Curiosity rover implies a cold and dry climate[42], although these deposits formed well after the valleys did[58]. It has also been suggested that chemical alteration within ancient lake deposits on Mars is considerably less than on Earth[90], but wind and volcanic deposits can obscure the evidence and make such orbital deductions inconclusive[11]. More telling is that opaline silica, a mineral associated with arid terrains on Earth, is altered to other minerals in wetter climates. Moreover, its presence alongside other alteration products in valley terrains suggests considerable alteration[11]. In addition, the presence of kaolin minerals in sedimentary sequences is typical of wet terrains on Earth and perhaps on early Mars[91]. Geochemical arguments that supposedly counter a warm and wet Mars are themselves contradicted by the large amounts of water that were clearly required to form the extensive valley networks[3] (Fig. 1), suggesting again that significant alteration must have occurred.

A hydrogen-rich atmosphere is possible because the early mantle appears to have been extremely reduced based on studies of Martian meteorities[40], although differing opinions exist[41]. Even though hydrogen is very light and can escape easily to space, hydrogen outgassing could have potentially outpaced any losses. Although the outgassing of high $CO_2$ concentrations may seem unlikely for a reduced mantle[42], $CO_2$ could have been produced indirectly due to reactions between atmospheric $H_2O$ vapor and outgassed reduced products, such as CO and $CH_4$[23,37,43].

**ICY HYPOTHESES**

Instead of a continuously warm climate maintained by atmospheric greenhouse gases, other scenarios in which an icy Mars is transiently warmed have been proposed. Some models involve impacts, including impact-induced steam atmospheres[44,45], runaway greenhouses[46], and cirrus clouds[47]. Another recent idea, the icy highlands hypothesis[26,42] also proposes that limited supplies of water (< 200 m GEL) were episodically released by transiently warming a cold Martian surface. This water would be stored and stabilized at high elevations in glacial ice sheets. The planet would undergo episodic warming, perhaps triggered through volcanism, impacts, or $CH_4$ bursts[24,26,44,48,49], triggering glacial melt generated from high elevations to carve the valley networks located at lower elevations.

Nevertheless, nearly all episodic warming mechanisms fail to generate the durations of warming and amounts of water required to form the modified craters or valleys. The peak of the purported Late Heavy Bombardment period (4.1 – 3.8 Ga) when the Earth is thought to have experienced a pulse of large impact events (Figure 3), occurred hundreds of millions of years before valley network formation[50]. Erosion rates are also thought to have been considerably higher during valley network formation than during the Late Heavy Bombardment[3,51], even though the valley formation period was characterized by subsequently smaller impactors[15,50]. This suggests that most, if not all, valley network formation had a non-impact origin. Impact-induced cirrus clouds provide some warming[47], but not enough[37,52] (see Box 2), and impact-induced runaway greenhouse atmospheres[46] would have quickly recondensed[52]. Even assuming regolith material that is relatively easy to erode (see Supp Info), predicted erosion rates from impact events[45] are still too low to have formed ancient landscapes by at least an order of magnitude[3]. Furthermore, calculated erosion rates from impact models[44,45], may be overestimated because they assume that all atmospheric water vapor immediately accumulates on the ground and does not escape to space. It has been suggested that erosion rates inferred from observations could themselves be overestimated because burial processes might have been dominant[53]. However, there are arguments that crater modification dominated by burial would result in different crater morphologies than what is observed[3] and computed burial volumes are dwarfed by erosion[3,54] (see Supp Info). Thus, it is unlikely that calculations that include sloping topography[53] would lead to significantly better agreement between modeled and observed erosion rates, especially since valley networks can traverse regions with relatively flat topography[3].

# BOX 2

> *TRANSIENT WARMING VIA CIRRUS CLOUDS*
>
> Cirrus clouds produce net warming on the Earth[92] and similarly may have warmed early Mars. Three-dimensional model simulations suggest that cirrus clouds composed of particles at least 10 µm in size could have sufficiently warmed early Mars to explain valley formation[47]. Existing models assume that the warming was dominated by water vapor and $CO_2$ was minimal, and simulate an initially cold planet with a frozen northern polar cap water reservoir that is mobilized by an external stimulus, such as obliquity variations, volcanism, or impacts to form global cirrus clouds. Given Mars' low gravity, both cloud particle fall times and cloud warming could have been greater than a similar scenario on Earth. However, just how much faster the cloud particles fall, which is related to the strength of greenhouse warming, is debated[47,52]. And even using the most ideal assumptions and 100% relative humidity, climate simulations suggest cirrus cloud cover of at least 75% would have been necessary to sufficiently warm early Mars[52]. Such high cirrus cloud fractions are inconsistent with what we know about cloud formation on Earth. Thus, unless downdrafts of drier air were unusually uncommon on early Mars, cirrus cloud warming by itself would not have produced above-freezing mean surface temperatures, even by 1) ignoring low clouds that would have cooled the climate further and, 2) including additional greenhouse warming from $CO_2$-$H_2$[37].

It has been hypothesized that serpentinization, a process by which heated groundwater interacts with iron- and magnesium-rich basaltic rocks at depth[55], periodically produced $CH_4$ or $H_2$ to warm the early Martian climate[24,55,56]. However, there is no evidence that this process has ever occurred on the massive scale required to produce enough $CH_4$ or $H_2$ to generate warm surface conditions[37] and how this process could have repeatedly occurred is unclear. Although impacts have been proposed as a means to destabilize the cryosphere and trigger massive $CH_4$ releases[56], they face the challenges explained above and in Box 2.

Chaotic obliquity transitions that destabilize methane trapped in the subsurface could produce episodic methane bursts[57], but at $CH_4$ concentrations insufficient to generate the long-lived warm climates apparently necessary for valley formation. More importantly, $CH_4$ is highly susceptible to destruction by solar ultraviolet radiation, which can remove several percent within ~ 100 – 250,000 years[24]. This timescale is sufficient to form smaller valleys at later times, and arguably Gale Crater deposits[58], but may only be a fraction of the time required to form the larger valley networks[59], unless discharge rates and sediment fluxes are very high[60]. Higher discharge rates would result in valley systems that are better integrated with the cratered landscape than what is actually observed[61], so valley formation likely occurred under clement atmospheric conditions characterized by relatively low discharge rates and longer formation timescales. Likewise, the hot and heavy rainfall predicted for some impact models[45,46] also suggest rapid (months – years) valley formation timescales[61] that contradict observations.

In some icy early Mars scenarios, liquid water is mostly limited to the subsurface and related to hydrothermal fluids, impact melts, or higher interior heat fluxes[62-65]. However, detailed geomorphic analyses have challenged these scenarios. The amount of water estimated to have formed the valleys is approximately 4,000 times the volume of the eroded valleys themselves, requiring repeated recycling of water through a stable hydrologic cycle[66]. Estimates suggest that runoff rates of ~10 cm/yr, averaged over 30 – 40 million years were necessary[59]. Aquifer recharge mechanisms in colder climates would be unable to effectively cycle such enormous water amounts[3,66]. Seasonal melting of snow in a cold climate would not have produced enough water to carve the valleys either[67]. Plus, the lack of mass wasting from freeze-thaw cycles, and the evidence for diffusional (rainfall) sediment transport in modified crater morphologies, both argue against snowmelt, which is exclusively an advective (surface runoff) process[3,54]. In addition, the presence of V-shaped valleys argues for precipitation and surface runoff because groundwater sapping would create flat-floored, U-shaped valleys instead[3,68].

According to the icy highlands hypothesis, snowmelt from icy highland areas was the main eroding agent. This implies that weathered landscapes would be localized to ice-covered areas and predicted glacial melt paths. This is not observed. The widespread absence of craters with a radius <5 km on ancient terrains[69], modification of older craters through both diffusional (rainfall) and advective (surface runoff) processes suggesting ubiquitous fluvial transport and erosion[54], and the widespread nature of fluvial features, such as those in Arabia Terria[70], are all consistent with a global process, likely rainfall, being the main agent of erosion on early Mars[3]. Moreover, melting from an extensive snowpack would permit valley networks to cross topographic divides and flow independently of the regional topography[70,71], and this is not observed[71,72]. Valleys also tend to grow in size downstream and be sourced from multiple contributing areas, which is atypical for melting snow[71,72]. Snowmelt flowing from higher elevations cannot explain the formation of valleys already located at high elevations, such as near crater rims[3]. In addition, the complete absence of glacial features (e.g., cirques, eskers, kames) in Noachian terrains, along with the lack of evidence for periglacial processes at Gale crater[58], do not support the idea that early Mars was heavily glaciated. Indeed, the estimated 5-km global equivalent layer (GEL) of water implies an ocean and an active hydrologic cycle, which is difficult to reconcile under such cold climates[66].

GCM simulations of the icy highlands hypothesis have been tuned such that they predict rainfall distributions that generally agree with the observed valley network distribution[75]. However, the simulations predict little rainfall in some valley network regions, such as Margaritifer Sinus, that exhibit characteristics consistent with rainfall[73,74]. In addition, the distribution of valley networks observed on modern day Mars does not include those networks that have been eroded or subsequently buried[76]. Thus, the icy highlands scenario[75] and other icy scenarios such as climate limit cycles in which extended periods of glaciation are punctuated by warm periods due to an active planetary carbon recycling mechanism[77] (see Supp Info), are not supported by the available geologic evidence. Not only are such limit cycles difficult to reconcile with a lack of evidence for widespread glaciation[57,70-72], but warm climates during valley network formation may have been stable against limit cycles[37].

In contrast to the geomorphic evidence, the geochemical evidence exhibits more ambiguity (also see Box 1). The observed sequence of Al-bearing clays being deposited over Mg-Fe-bearing clays in ancient terrains (e.g. NE Syrtis, Mawrth Vallis, Arabia Terria, Valles Marineris) has been inferred to be consistent with a largely frozen early Martian surface[62]. However, such sedimentary sequences are not necessarily indicative of a cold climate because rainwaters in a $CO_2$–rich early atmosphere supplemented by $SO_2$ volcanic outgassing would have been acidic. The observed geologic sequences are the expected byproducts of top-down weathering from acidic rain as Na, Mg, and Ca are leached from upper layers and Mg-Na-Ca sulfate-chloride solutions are precipitated at depth[78].

## THE ICE ALBEDO PROBLEM

An icy early Mars creates another problem: a glaciated planet requires higher greenhouse gas concentrations to warm it than does a less icy one[37]. This is because ice is reflective and reduces the amount of energy available to warm the surface (Figure 4). In the absence of such ice, the surface temperature in a representative cloud-free 1-bar atmosphere is 234 K (Figure 4a). Snow/ice mixtures that cover ~20% of the surface would reduce surface temperatures by 11 K (Figure 4b). The surface temperature decreases by 25 K if the ice coverage were doubled (40%), (Figure 4c). In this latter case, it is not possible to further increase the surface temperature by increasing greenhouse gas pressures (while keeping constant $H_2$ and $CO_2$ concentrations) because the surface is too icy (Figure 4c)[37]. This latter situation is equivalent to atmospheric collapse in 3-D models[22]. In the ice-free case, however (Figure 4a), warmer surface temperatures are still possible at even higher pressures, and the freezing point of water can conceivably be exceeded[37].

3-D simulations of nearly pure $CO_2$ atmospheres find that $CO_2$ clouds may partially mute the ice surface albedo effect[22]. However, if the atmosphere contains substantial $H_2$, this would be accompanied by significant $CH_4$[23,36,37], which heats the upper atmosphere and drastically inhibits $CO_2$ cloud formation[23,36,37]. Thus, the ice albedo effect would remain a significant problem for a $H_2$- or $CH_4$-rich cold Mars. Moreover, the high thermal inertia of ice may further exacerbate the problem[22].

Simulations of icy early Mars scenarios suggest that global ice coverage could have exceeded 25- 30% [26]. It is difficult to identify a transient warming mechanism that can repeatedly warm an already frozen surface[37]. The ice problem could be even greater than shown in Figure 4 if the surface is dominated by fresh snow, the ice is thick, or if ice coverage exceeds the 20 – 40% assumed here. Thus, a relatively arid planet that later underwent environmental changes that produced high surface runoff, triggering Martian valley formation, is consistent with the lack of evidence for icy features[3,5,6,79].

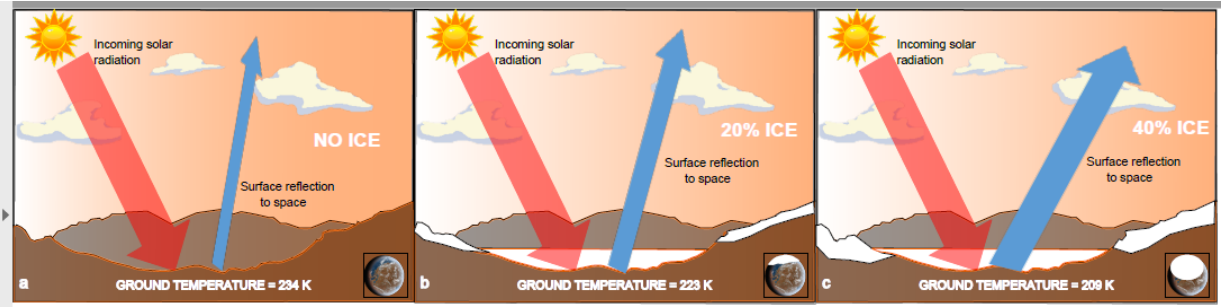

**Figure 4:** Illustration of simplified energy balance for Mars. Incoming (red arrow) and outgoing (blue arrow) solar energy required to support 1 bar $CO_2$ with 1% $H_2$ cold early Mars atmospheres with snow/ice (65% reflectivity) surface fractions of a) 0%, b) 20%, and c)40%. As surface snow/ice coverage grows, the amount of energy reflected back out to space increases (represented as a widening blue arrow), cooling the surface. Calculations based on ref[37].

## ARGUMENT FOR A WARM AND SEMI-ARID CLIMATE

Arguments for a cold and icy early Mars have persisted because of the difficulty in explaining the magnitude of greenhouse warming necessary to offset the faint young sun. However, geologic observations and climate models are beginning to merge[3,37]. The geologic evidence indicates that most of the earliest history of Mars, before valley network formation, was dominated by modification of impact craters through a combination of rain splash and limited surface runoff[79]. With time, more humid conditions were supported, rainfall intensified, and valley network formation was initiated (Figure 2)[5,6,13]. However, this "humid" early Mars may have been no warmer or wetter than the Great Basin region of North America during the Pleistocene glaciation[13], where seasonal warming supported rainfall and surface runoff. Such a semi-arid climate is consistent with the lower end of global water estimates (< 200 m GEL)[9]. On Mars, rainfall and surface runoff may have been enhanced by lower atmospheric pressures of < 2 bars[79]. Unlike a perennially cold and icy highlands scenario[26], thin ice or snow accumulations would seasonally melt as the freezing point was reached, reducing the surface albedo and greatly diminishing the ice problem. This would explain the lack of evidence for extensive wet-based glaciation after the end of the warm period[42]. Alternatively, perhaps Mars was even warmer and maybe even persistently so, but water was simply too limited to generate the higher erosion rates associated with equally warm climates on the Earth. Most of the planet may have been very arid and cold, with precipitation rates orders of magnitude lower than the regions of valley network formation where conditions were more semi-arid. By analogy, the driest regions of the Atacama Desert on Earth are orders of magnitude drier than terrestrial semi-arid terrains[80], which are orders of magnitude drier than tropical regions. The zones of relatively higher precipitation on Mars could have shifted over time and location, mirroring spatial rainfall variations on the Earth[81]. Nevertheless, although snow and snowmelt were likely part of the history of Mars, the geomorphology cannot be explained without the occurrence of above-freezing surface temperatures and rain (seasonally – if not persistently). The frequent occurrence of above-freezing temperatures potentially avoids the need for high temperatures (~25 – 50 C) and multi-bar atmospheres (>~ 5 – 10 bar) to form the observed distribution of surface materials[37,82], which could be explained by a volcanically-produced $CO_2$-$H_2$ and/or a $CO_2$-$CH_4$ greenhouse atmosphere[37]. Thus, we suggest that the geologic evidence can be reconciled with climate model predictions.

The valley networks seem to have been carved by water, but neither a multi-bar (> a few bar) early atmosphere or transient warming scenarios that invoke icy climates fit the existing observational evidence. Scenarios that advocate very short (months – years) valley formation timescales are also disfavored. The geologic evidence suggests that at least seasonal – if not persistent- episodes of warmth and precipitation were required to produce the observed geomorphology. The most ancient Noachian terrains that predate valley network formation are heavily-cratered and complicated[5]. Nevertheless, observations indicate that there was a general lack of glaciation, relatively low erosion rates, and minimal fluvial dissection[5,6]. All of this suggests an arid Noachian climate before fluvial erosion increased, leading to valley network formation[3,5-6,79]. We hypothesize that the climate on early Mars was arid during the Noachian and became semi-arid around the Noachian-Hesperian boundary (Figure 3) with the climatic shift due to warming by volcanic emissions of greenhouse gases. More geomorphic analyses and climate modeling, especially of the period preceding valley network formation, are needed to test this hypothesis.

Corresponding author: Ramses Ramirez, email: rramirez@elsi.jp

## ACKNOWLEDGEMENTS

R.M.R. wishes to thank James F. Kasting for lively discussions about limit cycles and transient warming episodes. R.M.R. acknowledges support from the Simons Foundation (SCOL # 290357, Kaltenegger), Carl Sagan Institute, Cornell Center for Astrophysics and Planetary Science, and the Earth-Life Science Institute. R.A.C. acknowledges support from NASA grant 80NSSC17K0454 and a grant from the Smithsonian's Universe Consortium.

There were no competing financial interests in the production of this work.

# AUTHOR CONTRIBUTIONS

R.M.R conceived idea, wrote and edited much of the main text and Supp. Info. and created the figures. R.A.C. defined the initial conditions of the geologic environment before valley network formation (early-mid Noachian) and co-wrote and co-edited the main text and Supp. Info. Both authors discussed and analyzed the results and implications.